\documentclass{raa}
\usepackage{graphicx,times}
\usepackage{natbib}
\usepackage{amssymb,amsmath}
\bibpunct{(}{)}{;}{a}{}{,}
\usepackage[colorlinks=true,linkcolor=green,citecolor = blue]{hyperref}%
\makeatletter
\newcommand*{\rom}[1]{\expandafter\@slowromancap\romannumeral #1@}
\makeatother

\def\fs{\hbox{$.\!\!^s$}}
\def\degr{\hbox{$^\circ$}}

\begin{document}%

\title{Modified masses and parallaxes of close binary system: HD\,39438}

\volnopage{ {\bf 2023} Vol.\ {\bf 00} No. {\bf 0}, 000--000}
   \setcounter{page}{1}

   \author{Suhail Masda\inst{1,2}, Z. T. Yousef\inst{3}, Mashhoor Al-Wardat\inst{4,5}, and Awni Al-Khasawneh\inst{4}}

\institute{Department of Physics, Mahrah University, Mahrah, Yemen; \textit{suhail.masda@gmail.com} \\
	     \and Department of Physics, Hadhramout University, Mukalla, Yemen\\
	      \and Department of Applied Physics, Faculty of Science and Technology, Universiti Kebangsaan Malaysia, 43600 UKM Bangi, Selangor, Malaysia\\
	      \and Department of Applied Physics and Astronomy, and Sharjah Academy for Astronomy, Space Sciences and Technology, University of Sharjah, P.O.Box 27272 Sharjah, United Arab Emirates\\
	       \and Department of Physics, Al al-Bayt University, Mafraq 25113, Jordan\\
\vs \no
   {\small Received xxxx Month xx; accepted yyyy Month yy}
}

\abstract{
We present the detailed fundamental stellar parameters of the close visual binary system; HD\,39438 for the first time. We used Al-Wardat's method for analyzing binary and multiple stellar systems (BMSSs). The method implements Kurucz's plane parallel model atmospheres to construct synthetic spectral energy distributions for both components of the system. It then combines the
results of the spectroscopic analysis with the photometric analysis and then compares them
with the observed ones to construct the best synthetic spectral energy distributions
for the combined system. The analysis gives the precise fundamental parameters of the individual
components of the system. Based on the positions of the components of HD\,39438 on the H-R
diagram, and evolutionary and isochrones tracks, we found that the system belongs to the main
sequence stars with masses of 1.24 and 0.98 solar masses for the components A and B, respectively,
and age of 1.995 Gyr for both components. The main result of HD\,39438 is new dynamical parallax, which is estimated to be $16.689\pm0.03$ mas. 
}

\authorrunning{Masda and Yousef}            
   \titlerunning{Fundamental stellar parameters of HD\,39438}  
\maketitle%

\section{INTRODUCTION }
One of the most vital disciplines for contemporary stellar astronomy is that is reliable and bound up with the understanding of the binary stars. The close binary systems are one of those binaries which make use of them in estimating the fundamental stellar parameters especially stellar masses precisely because the majority of systems makes up 50\% from the binary and multiple systems \citep{1991A&AS...88..281D}.

The techniques of speckle interferometry \citep{2002AA...385...87B,2010AJ....139..743T,2014CoSka..43..229K} and adaptive
optics \citep{2005AJ....130.2262R,2011MNRAS.413.1200R}, which are modern techniques, are instrumental in studying and analyzing  the close visual binary systems. The analysis of spectroscopic  and astrometric data is of overriding significance in the period of the speckle
interferometry and adaptive optics \cite{2018A&A...618A.100L}. These techniques are also instrumental in improving the situation for those binaries in terms of evolutionary status and relative motion.

Al-Wardat's method for analyzing binary and multiple stellar systems (BMSSs) involves merging the results of  spectroscopic and photometric analyses, with the observed measurements. This combination is regarded as the most effective method for analyzing such systems \citep{2002BSAO...53...51A,2003BSAO...55...18A,2007AN....328...63A}. It merges the magnitude difference measurements of the speckle interferometry, the combined spectral energy distributions (SEDs) of the spectrophotometric analysis with the aid of the grids of Atlas9 models \cite{1994KurCD..19.....K}, and radial velocity measurements (once available) to estimate the individual fundamental stellar parameters of the binary systems, thereby determining the precise spectrophotometric masses of the binary systems. The method has been utilized to compare the synthetic stellar photometry with the observed stellar photometry to estimate the fundamental parameters of the solar-type stars  ~\citep{2009AN....330..385A,2012PASA...29..523A,2014AstBu..69...58A,2014AstBu..69..198A,2016RAA....16..112M,2016RAA....16..166A,2017AstBu..72...24A,2018RAA....18...72M,2018JApA...39...58M,2019RAA....19..105M,2019AstBu..74..464M,2021RAA....21..161A,2021AIPC.2335i0002M,2021RAA....21..110W,2021RAA....21..114Y,2014PASA...31....5A,2014AstBu..69...58A,2014AstBu..69..454A}. 

On the one hand, the synthetic stellar photometry is mainly utilized to obtain the fundamental stellar parameters more precisely through comparing their results with the observed ones \citep{2014AstBu..69...58A,2014AstBu..69..198A,2018RAA....18...72M,2018JApA...39...58M,2019RAA....19..105M,2019AstBu..74..464M,2021RAA....21..161A,2021AIPC.2335i0002M,2021RAA....21..110W}. The synthetic photometry is synthetic analysis of the results of the spectroscopic analysis (synthetic SED) of the binary system. This contributes to the improvement the fundamental  parameters of the close binary systems. 

In this study, the studied system of HD\,39438 (HIP\,27758) is a visual close binary in solar neighborhood located at $\pi_{\rm H07}=20.15\pm1.19$ mas \cite{2007A&A...474..653V} and at $\pi_{\rm DR3}=16.0508\pm0.264$ mas \cite{2022yCat.1355....0G}, where the abbreviations of the astrometric measurements have been pointed out in \cite{MASDA2023}. Since, the renormalized unit weight error (RUWE) in $\pi_{\rm DR3}$ was very large, the solution of this parallax should not be used in that case.

The first orbit for this binary was solved by \cite{2010AJ....140..735M} with grade of three (G 3 = Reliable). Acoording to \cite{2010AJ....140..735M}, they found that the dynamical stellar mass was $\mathcal{M} =1.29\pm0.44\mathcal{M}_{\odot}$ based on  H07's parallax, and $2.56\pm0.34\mathcal{M}_{\odot}$ based on the DR3's parallax. Then, the second orbit was studied by \cite{2014AJ....147..123T} with grade of three and the dynamical stellar mass of the system was $\mathcal{M} =1.53\pm0.28\mathcal{M}_{\odot}$ based on H07's parallax, and $\mathcal{M} =3.02\pm1.70\mathcal{M}_{\odot}$ based on DR3's parallax. In 2017, \cite{2017AJ....154..110T} revised the orbit of the system. In that case, \cite{2017AJ....154..110T} found that the orbit has been graded as 2 (G 2=Good), which should be adopted. According to this study, the individual dynamical masses were estimated to be $\mathcal{M} =1.29\mathcal{M}_{\odot}$ and $\mathcal{M} =0.97\mathcal{M}_{\odot}$ for the primary and secondary components, respectively. In this solution, the dynamical stellar mass of the system was $\mathcal{M} =1.50\pm0.27\mathcal{M}_{\odot}$ based on H07's parallax, and $\mathcal{M} =2.97\pm0.15\mathcal{M}_{\odot}$ based on DR3's parallax. Based on these results, \cite{2017AJ....154..110T} indicated that the H07's parallax is not accurate  and should be revised more precisely. As a result, \cite{2017AJ....154..110T} suggested new dynamical parallax as $\pi_{\rm dyn}=17.6$ mas.

In this study, we proceed with the series of \cite{MASDA2023}, which presents the modified masses and parallaxes for a selected sample  of close binary systems. First, the key aim for this paper is to present the fundamental stellar parameters of the binary system utilizing  Al-Wardat's method for analyzing BMSSs. Second, we present the comparison  between the observed photometry of the combined system with the combined synthetic photometry. Third, we present the spectrophotometric stellar masses and new dynamical parallax of the system.

In this studt, we present the observational data of HD\,39438 (HIP\,27758), which will be compared to the synthetic analysis in section \ref{section2}. Section \ref{section3} containes the analysis method of the binary systems. Section \ref{section4} gives the method to calculate the stellar masses of the binary system. In section \ref{section5}, the results and discussion of the close binary system present. Finally, in section \ref{section6}, we present the conclusion of this study.

\section{Observed data}\label{section2}
The observed data is the backbone of the synthetic analysis. The observed photometric data, which is  available from miscellaneous sources, is the key refernces to obtain the stellar parameters. These are Hipparcos catalogue \citep{1997yCat.1239....0E}, Str\"{o}mgren \citep{1998A&AS..129..431H} and Tycho 
catalogues \citep{2000A&A...355L..27H}, which will be compared with the synthetic photometric data to study details of the system. Table~\ref{t1} contains the basic data and the observed photometric data of HD\,39438, while Table~\ref{t2} contains the observed magnitude differences between the primary and secondary components of the binary system.

\begin{table*}
	\centering
	\caption{Fundamental data and observed  photometry of HD\,39438}\label{t1}
	\begin{tabular}{ccc}
		\hline\hline
		Property & HD\,39438 &  Ref.  \\
		& HIP\,27758 &   \\
		\hline
		$\alpha_{2000}$   & $05^{\rm h} 52^{\rm m} 29\fs411$ &    1\\
		$\delta_{2000}$  & $-02\degr17' 07.''62$ &  1\\
		Sp. Typ.  &  G0V &   2
		\\
		Gaia  DR2 &  3025640770239673216 &  1\\
		Gaia  DR3 &  3025640770239673216 &  1
		\\
		$E(B-V)$& $ 0.017 $ & $3$
		\\		
		$A_{\rm v}$ (mag)  &  $0.053$ & $*$
		\\
		$\pi_{\rm H07}$ (mas) &    $20.15\pm1.19$   &  $4$
		\\
		$\pi_{\rm DR2}$  (mas) &    $11.906\pm0.37$  & $5$\\
		$\pi_{\rm DR3}$  (mas) &    $16.051\pm0.26$  & $5$\\
		$[\rm Fe/H]$  & $-0.12\pm0.08$ &   $ 6 $ \\
		$V_{\rm J}$ (mag) & $7.26$  &  $7$
		\\
		$B_{\rm J}$ (mag)  &  $7.79\pm0.02$   & 8
		\\		
		$(B-V)_{\rm J}$ (mag)& $0.56\pm0.015$ & 7
		\\
		$(b-y)_{\rm S}$ (mag)& $0.35$ &  $9$
		\\
		$(v-b)_{\rm S}$ (mag)& $0.51$  & 9\\
		$(u-v)_{\rm S}$ (mag)& $0.88$ & 9\\
		$B_{\rm T}$ (mag) &   $7.89\pm0.012$  & $7$
		\\	
		$V_{\rm T}$ (mag) &   $7.32\pm0.010$    & 7
		\\
		\hline
	\end{tabular}\\
	\textbf{Notes.} $*$ means $A_{\rm v}=3.1 E(B-V)$, Ref. $^1$ \citep{2020yCat.1350....0G}, $^2$ \citep{1999MSS...C05....0H}, $^{3}$ \citep{2014A&A...561A..91L}, $^4$ H07's parallax \citep{2007A&A...474..653V}, $^5$ DR2 and DR3's parallax \citep{2016A&A...595A...1G,2022yCat.1355....0G}, $^6$ \citep{2016ApJ...826..171G}, $^7$ \citep{1997yCat.1239....0E},
	$^{8}$ \citep{2000A&A...355L..27H}, and $^{9}$ \citep{1998A&AS..129..431H}.
\end{table*}

\begin{table}[h]
	\begin{center}
		\caption{The observed magnitude difference between the components of HD\,39438 (HIP\,27758).}
		\label{t2}
		\begin{tabular}{c|cccc}
			\noalign{\smallskip}
			\hline\hline
			HD	&$\triangle m $& {$\sigma_{\Delta m}$}& Filter ($\lambda/\Delta\lambda$)& Ref.  \\
			\hline
			39438	&	$0.87$ &   0.89  & $V_H:511$nm/222&  \cite{1997yCat.1239....0E}  \\
			&	$1.34$&  $ 0.03$  & $545$nm/30&  \cite{2005AA...431..587P}  \\
			&	$1.31$ &   0.03  & $545$nm/30&  \cite{2002AL...28..773B}   \\
			&	$1.65$ &   & $550$nm/40&  \cite{2008AJ....136..312H}   \\
			&	$1.26$ &   & $541$nm/88&  \cite{2008AJ....136..312H}  \\
			&	$1.33$ &   & $550$nm/40& \cite{2010AJ....139..205H}   \\
			&	$1.40$ &   & $551$nm/22&  \cite{2010AJ....139..743T}   \\
			&	$2.30$ &   & $543$nm/22&   \cite{2012AJ....143...42H} \\
			&	$1.60$ &   & $543$nm/22& \cite{2014AJ....147..123T}   \\
			&	$1.50$ &   & $543$nm/22&  \cite{2014AJ....147..123T}   \\
			&	$1.60$ &   & $543$nm/22&  \cite{2015AJ....150...50T}  \\
			&	$1.62$ &  $0.28$ & $543$nm/22&  \cite{2017AJ....154..110T}   \\
			&	$1.60$ &   & $543$nm/22&  \cite{2016AJ....152..138T}   \\
			\hline
		\end{tabular}
	\end{center}
\end{table}

\section{Method and Analysis}\label{section3}
The spectrophotometric analysis is the most important step in Al-Wardat's method, which depends on two solutions to estimate the astrophysical parameters. The solutions are as follows:
\subsection{Spectroscopic solution}\label{22}
The spectroscopic solution is the most important key to reach the fundamental stellar parameters. In this solution, we require to construct the synthetic SED for the combined and individual synthetic SED of the system. First of all, we need to know the observed magnitude difference of the system, which is estimated as follows: $\triangle m=1.49\pm0.01$. This parameter is the average for all $\triangle m$ measurements given in Table~\ref{t2} under the V-band filters. The observed magnitude difference of the system, combined with the visual magnitude, led to the individual apparent and absolute magnitudes  of the system as follows: $ m_v^A=8^{\rm m}.30\pm0.002$, $\rm M_V^A=4^{\rm m}.36\pm0.01$, and $ m_v^B=8^{\rm m}.64\pm0.12$, $\rm M_V^B=4^{\rm m}.70\pm0.13$ for the primary and secondary components, respectively, by using the following simple relationships:
\begin{eqnarray}\centering
\label{eq1}
\rm m_v^A=\rm m_v+2.5\log(1+10^{-0.4\triangle\rm m}),
\end{eqnarray}
\begin{eqnarray}
\centering
\label{eq2}
\rm m_v^B=\rm m_v^A+{\triangle\rm m},
\end{eqnarray}
\begin{eqnarray}
\label{eq3}
\rm M_V-m_v=5-5\log(d)-A_V,
\end{eqnarray}
 Here, the distance of the system from Earth ($d$) is measured in parsec (pc). Furthermore, since HD\,39438 is a nearby system, the interstellar extinction is neglected.  
 
The absolute magnitudes of HD\,39438 components are employed for estimating the input parameters, together with some parameters taken as introductory values from the Tables of \cite{1992adps.book.....L} and \cite{2005oasp.book.....G}. In addition, the following equations for the main sequence stars are used:
\begin{eqnarray}
\label{eq4}
\log\frac{R}{R_\odot}= \frac{M_{bol}^\odot-M_{bol}}{5}-2\log\frac{T_{\rm eff}}{T_\odot},\\
\label{eq5}
\log g = \log\frac{M}{M_\odot}- 2\log\frac{R}{R_\odot} + \log g_\odot.
\end{eqnarray}	
\noindent where $T_\odot=5777\,\rm K$, log $ g_\odot=4.44$ and $M_{bol}^\odot=4^{\rm m}.75$. $M_{bol}=\rm M_{V}+BC$; where $\rm BC$ is the bolometric correction.

The individual synthetic SED for each single star are built based on input parameters of the binary system. Therefore, the Kurucz Atlas9 models, which are plaane-parallel model atmospheres developed by Kurucz in 1994, are empployed. These models are used to generate the synthetic fluxes for individual components of the system, and when combined with the parallax, they produce the synthetic SED for the combined close binary system. For this purpose, the specialized subroutines of  Al-Wardat's method for analyzing BMSSs must be utilized. The combined synthetic SED of the binary system is determined using the following equation:
\begin{eqnarray}
\label{eq6}
F_{\lambda, s}  = \Bigg(\frac{R_{A}}{d}\Bigg)^2\Bigg(H_\lambda ^A + H_\lambda ^B \Bigg(\frac{R_{B}}{R_{A}}\Bigg)^2\Bigg)
\end{eqnarray}
\noindent
where $F_\lambda$ is the combined synthetic SED of the binary system, $ R_{A}$ and $ R_{B}$ are the radii of component A and component B  of the system, respectively, in solar units. 
 $H_\lambda ^A $ and  $H_\lambda ^B$ are the corresponding fluxes of component A and component B, respectively, in units of ergs cm$^{-2}$\ s$^{-1}$ \AA$^{-1}$. These individual fluxes are dependent on the $T_{\rm eff.}$ and log g. This equation accounts for the energy fluxx of the individual components located at a distance d (in parsecs) from Earth, ensuring a reliable estimation.

In Equation~\ref{eq6}, the values of the radii are  dependent mainly on the accuracy of the parallax measurements. \citep{Tokovinin2000} showed that the parallax measurements of the binary systems were probably distorted by the orbital motion. That is why, Al-Wardat's method discovered that the problems started appearing significantly in the parallax measurements of the binary systems and gives the new dynamical parallax \citep{2021PASA...38....2A,MASDA2023}.  

The results of the fundametal stellar parameters should be in keeping with those observed ones of the binary system. This has ranked as one of the best ways to make certain accuracy of the parallax of the system. So, synthetic photometric solution should be executed to determine the best stellar parameters of the close visual binary system.

\subsection{Photometric solution}
The synthetic photometric solution is the perfect complement to the spectroscopic solution, which is in turn instrumental in estimating  the fundamental stellar parameters of the close binary systems. The main aim of this solution is to calculate the magnitudes and color indices of the combined and individual synthetic SEDs and then compare with the observed ones  in any photometric system. So, the synthetic magnitudes and color indices in different photometrical systems such as: Johnson: $U$, $B$, $ V$, $R$, $U-B$, $B-V$, $V-R$; Str\"{o}mgren: $u$, $v$, $b$,
$y$, $u-v$, $v-b$, $b-y$ and Tycho: $B_{T}$, $ V_{T}$, $B_{T}-V_{T}$ are calculated by using the following equation \citep{2012PASA...29..523A}:
\begin{equation}\label{eq7}
m_p[F_{\lambda,s}(\lambda)] = -2.5 \log \frac{\int P_{p}(\lambda)F_{\lambda,s}(\lambda)\lambda{\rm d}\lambda}{\int P_{p}(\lambda)F_{\lambda,r}(\lambda)\lambda{\rm d}\lambda}+ {\rm ZP}_p\,
\end{equation}
\noindent	where $m_p$ is the synthetic magnitude of the passband $p$, $P_p(\lambda)$ is the dimensionless sensitivity function of the passband $p$, $F_{\lambda,s}(\lambda)$ is the synthetic SED of the object and $F_{\lambda,r}(\lambda)$ is the SED of the reference star (Vega).  Zero points (ZP$_p$) from~\cite{2007ASPC..364..227M} are adopted.

\section{Mass and dynamical parallax} \label{section4}

The stellar mass played a vital role in understanding the formation and evolution of the binary systems. Thus, its estimation should be accurate. There are two types of masses, which are the spectrophotometric mass $\mathcal{M}_{Sph}$ and dynamical stellar mass $\mathcal{M}_{d}$. The former is estimated based on the evolutionary tracks by using Al-Wardat's method for analyzing BMSSs, while the later is estimated by using orbital solution of the system based on Kepler's third law as follows:
\begin{eqnarray}
\label{eq8}
\mathcal{M}_{d}=\mathcal{M}_{A}+\mathcal{M}_{B}=\Big(\frac{a^3}{\pi^3P^2}\Big)\ \mathcal{M}_\odot,
\end{eqnarray}
The  error in the dynamical mass is estimated as follows:
\begin{eqnarray} 
\label{eq9}
\frac{\sigma_\mathcal{M} }{\mathcal{M}} =\sqrt{9\Big(\frac{\sigma_\pi}{\pi}\Big)^2+9\Big(\frac{\sigma_a}{a}\Big)^2+4\Big(\frac{\sigma_p}{p}\Big)^2}
\end{eqnarray}

where $a^{''}$ and $\pi$ are the semi-major axis and the parallax
(both in arcsec), respectively, P is the orbital period (in years), $\ \mathcal{M}_{A}$ and $\ \mathcal{M}_{B}$ are the masses (in
solar mass).

The  dynamical masses are dependent mainly on the grades of the orbits. We can adopt the best orbit if that orbit has the grades of  Grade 1=Definitive, Grade 2=Good, and Grade 3=Reliable. When the spectrophotometric mass is in keeping with the dynamical mass, the parallax of the system is adopted, otherwise it should be estimated by using Al-Wardat's method as follows:
\begin{eqnarray}
\label{eq10}
\mathcal{\pi}_{dyn}=\frac{a}{P^{2/3}(\sum \mathcal{M}_{Sph} )^{1/3}}
\end{eqnarray}
where $\sum \mathcal{M}_{Sph}$ are estimated by using Al-Wardat's method for analyzing BMSSs in solar mass and $\pi_{dyn}$ is in arcsec. Its error is estimated as follows:
\begin{eqnarray} 
\label{eq321}
\frac{\sigma_{\mathcal{\pi}_{dyn}}}{\mathcal{\pi}_{dyn}} =\sqrt{\frac{4}{9}\Big(\frac{\sigma_P}{P}\Big)^2+\Big(\frac{\sigma_a}{a}\Big)^2+
	\frac{1}{9}\Big(\frac{\sigma_{\sum \mathcal{M}_{Sph}}}{\sum \mathcal{M}_{Sph}}\Big)^2}
\end{eqnarray}

\section {Results and discussions}\label{section5}

The fundamental stellar properties  of the close binary system, HD\,39438 were estimated using the complex analytical method (Al-Wardat's method for analyzing BMSSs) by \citep{2002BSAO...53...51A}. The method combines the spectroscopic solution with photometric solution to estimate the physical and geometrical stellar parameters of the system.  These led to presenting a new value for the system`s parallax of HD\,39438.

The results of the calculated synthetic magnitudes and colour  indices of the individual components and combined synthetic SEDs of the binary system, HD\,39438 are listed in Table \ref{t3}. These have presented in different photometrical systems (Johnson: $U$, $B$, $ V$, $R$, $U-B$, $B-V$, $V-R$; Str\"{o}mgren: $u$, $v$, $b$,
$y$, $u-v$, $v-b$, $b-y$ and Tycho: $B_{T}$, $ V_{T}$, $B_{T}-V_{T}$). 

\begin{table}[ht]
	\small
	\begin{center}
		\caption{ The synthetic stellar photometry of  HD\,39438.}
		\label{t3}
		\begin{tabular}{ccccc}
			\noalign{\smallskip}
			\hline\hline
			\noalign{\smallskip}
			Sys. & Filter &  Combined Synth. & HD\,39438 & HD\,39438\\
			&     & $\sigma=\pm0.03$&   A    &     B      \\
			\hline
			\noalign{\smallskip}
			Joh-          & $U$ & 7.90 & 8.06 & 10.06 \\
			Cou.          & $B$ & 7.82   &  8.02 &  9.76  \\
			& $V$ & 7.26 &  7.51 &  9.00 \\
			& $R$ & 6.95  &  7.22 & 8.59  \\
			&$U-B$& 0.08  & 0.04 & 0.31\\
			&$B-V$& 0.56  &  0.51 &  0.76 \\
			&$V-R$& 0.31  &  0.29 & 0.40 \\
			\hline
			\noalign{\smallskip}
			Str\"{o}m.    & $u$ & 9.07 & 9.23 &  11.21  \\
			& $v$ & 8.14 & 8.32  & 10.16  \\
			& $b$ & 7.58 & 7.80 &  9.41 \\
			&  $y$& 7.23 & 7.48 & 8.96 \\
			&$u-v$& 0.93 & 0.91 & 1.05 \\
			&$v-b$& 0.56 & 0.52 & 0.675\\
			&$b-y$& 0.35 & 0.32& 0.45 \\
			\hline
			\noalign{\smallskip}
			Tycho       &$B_T$  & 7.96   & 8.14 & 9.95   \\
			&$V_T$  & 7.33   & 7.57& 9.08  \\
			&$B_T-V_T$& 0.63 & 0.57 & 0.87\\
			\hline\hline
			\noalign{\smallskip}
		\end{tabular}
	\end{center}
\end{table}

Table~\ref{t4} shows the best agreement between the synthetic and observed photometry of the binary system, HD\,39438. This agreement demonstrates that the basic stellar characteristics of each element in the system listed in  Table~\ref{t5} are reliable. Table~\ref{t3} indicates that the synthetic apparent magnitudes are completely in keeping with the observed apparent magnitudes of the system. 

\begin{table}[h]
	\small
	\begin{center}
		\caption{The best agreement between the observed photometry from catalogues and the synthetic photometry from this study of HD\,39438.} \label{t4}
		\begin{tabular}{ccc}
			\noalign{\smallskip}
			\hline\hline
			\noalign{\smallskip}
			&\multicolumn{2}{c}{HD\,39438}\\
			\cline{2-3}
			\noalign{\smallskip}
			Filter	& Observed $^a$ & Synthetic$^b$(This work) \\
			& ($\rm mag$) & ($\rm mag$) \\
			
			\hline
			\noalign{\smallskip}
			$V_{J}$ & $7.26$ & $7.26\pm0.03$\\
			$B_J$& $7.79\pm0.02$ & $7.82\pm0.03$  \\
			$B_T$  & $7.93\pm0.012$   &$7.96\pm0.03$\\
			$V_T$  & $7.32\pm0.01$   &$7.33\pm0.03$\\
			$(B-V)_{J}$&$ 0.56\pm0.02$ &$ 0.56\pm0.03$\\
			$(u-v)_{S}$&$ 0.88$ &$ 0.93\pm0.03$\\
			$(v-b)_{S}$&$ 0.51$ &$ 0.56\pm0.03$\\
			$(b-y)_{S}$&$ 0.35$ &$ 0.35\pm0.03$\\
			$\triangle\rm m$  &$ 1.49^{c}\pm0.01$  &$ 1.49^{d}\pm0.05$\\
			\hline\hline \noalign{\smallskip}
		\end{tabular}\\
		Notes:
		$^a$ The observational data of HD\,39438 (see Table~\ref{t1}), 
		$^b$ The synthetic photometry of HD\,39438 (see Table~\ref{t3}),
		$^c$ The observed magnitude difference of HD\,39438 (see Table~\ref{t2}) and 
		$^d$ The synthetic magnitude difference of HD\,39438 (see Table~\ref{t3}) 
	\end{center}
\end{table}

 The stellar luminosities of the individual components of HD\,39438  are estimated to be as follows: $L_A=2.65\pm0.08\,\rm L_\odot$ and $L_B=0.76\pm0.09\,\rm L_\odot$ for the primary and secondary omponents of the system, while the spectral types for them are  F5.5V and G8V, respectively, which are in line with the spectral types of \cite{2010AJ....140..735M} and \cite{2017AJ....154..110T}, and with the spectral type F5V given in WDS and SIMBAD catalogues.

\begin{table*}[h]
	\small
	\begin{center}
		\caption{The fundamental stellar parameters of the individual components of HD\,39438.}
		\label{t5}
		\begin{tabular}{cccc}
			\noalign{\smallskip}
			\hline\hline
			&       	&\multicolumn{2}{c}{H\,39438}  \\
			\cline{3-4}
			\noalign{\smallskip}
			Parameters & Units	& H\,39438 & HD\,39438  \\
				 & 	& A & B  \\
				\hline
				\noalign{\smallskip}
			$\rm T_{\rm eff}$ & [~K~] & $6370\pm100$ & $5580\pm100$ \\
			R   & [R$_{\odot}$] & $1.34\pm0.09$ & $0.935\pm0.09$\\
			$\log\rm g$  & [cgs] & $4.35\pm0.07$ & $4.50\pm0.07$ \\
			$\rm L $  & [$\rm L_\odot$] & $2.65\pm0.08 $  & $0.76\pm0.09$ \\
			$\rm M_{bol}$  & [mag] &  $3.69\pm0.08$ & $5.05\pm0.09$ \\
			$\rm M_{V}$  & [mag] & $4.03\pm0.12$ & $5.52\pm0.13$ \\
			M $^{a}$ & [$\rm M_{\odot}$]&  $1.24 \pm0.08$ & $0.98 \pm0.07$\\
			Sp. Type &   & F5.5V & G8V   \\
			\hline
			\multicolumn{1}{c}{Parallax $^{b}$  }
			& [mas]&  \multicolumn{2}{c}{$20.15 \pm 1.19 $}\\				
			\multicolumn{1}{c}{Age $^{c}$}
			& [Gyr]& \multicolumn{2}{c}{ $1.995$}\\
			\hline\hline
			\noalign{\smallskip}
		\end{tabular}\\
		$^{a}${Based on Al-Wardat's method}\\ 
		$^{b}${Based on H07's parallax}.\\
		$^{c}${Based on the the isochrones tracks}.	
	\end{center}
\end{table*}

\begin{figure}[ht]
	\centering
	\includegraphics[angle=0,width=12cm]{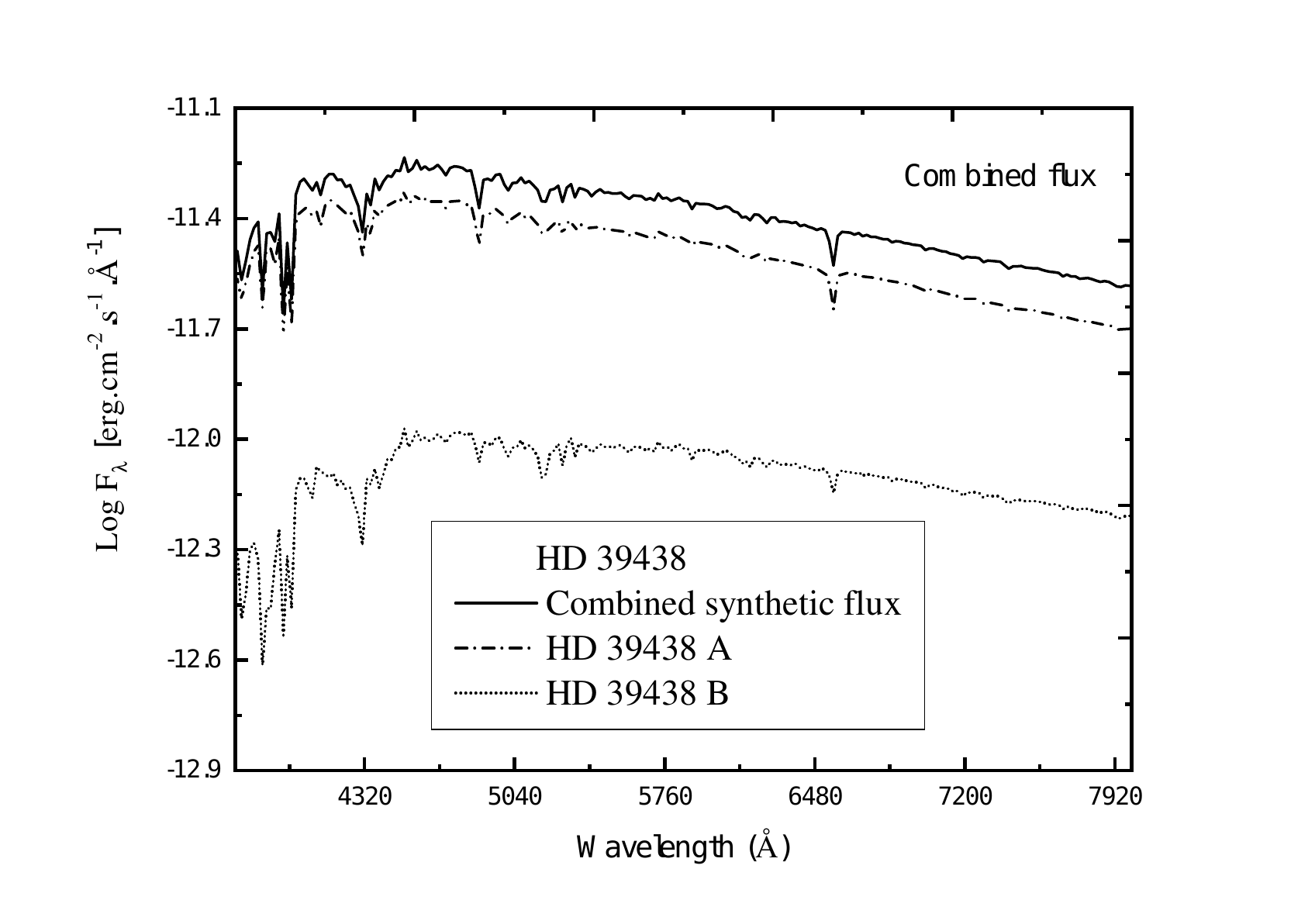}
	\caption{The combined synthetic SED and its individual components of HD\,39438.}
	\label{f1}
\end{figure}

 According to the results of the analysis, Fig.~\ref{f1} shows the adopted combined synthetic SED and  the synthetic SED for the individual component of the binary system for the first time based on the best agreement between the observed and synthetic stellar photometry of the system.

\begin{figure}[ht]
	\centering
	\includegraphics[angle=0,width=12cm]{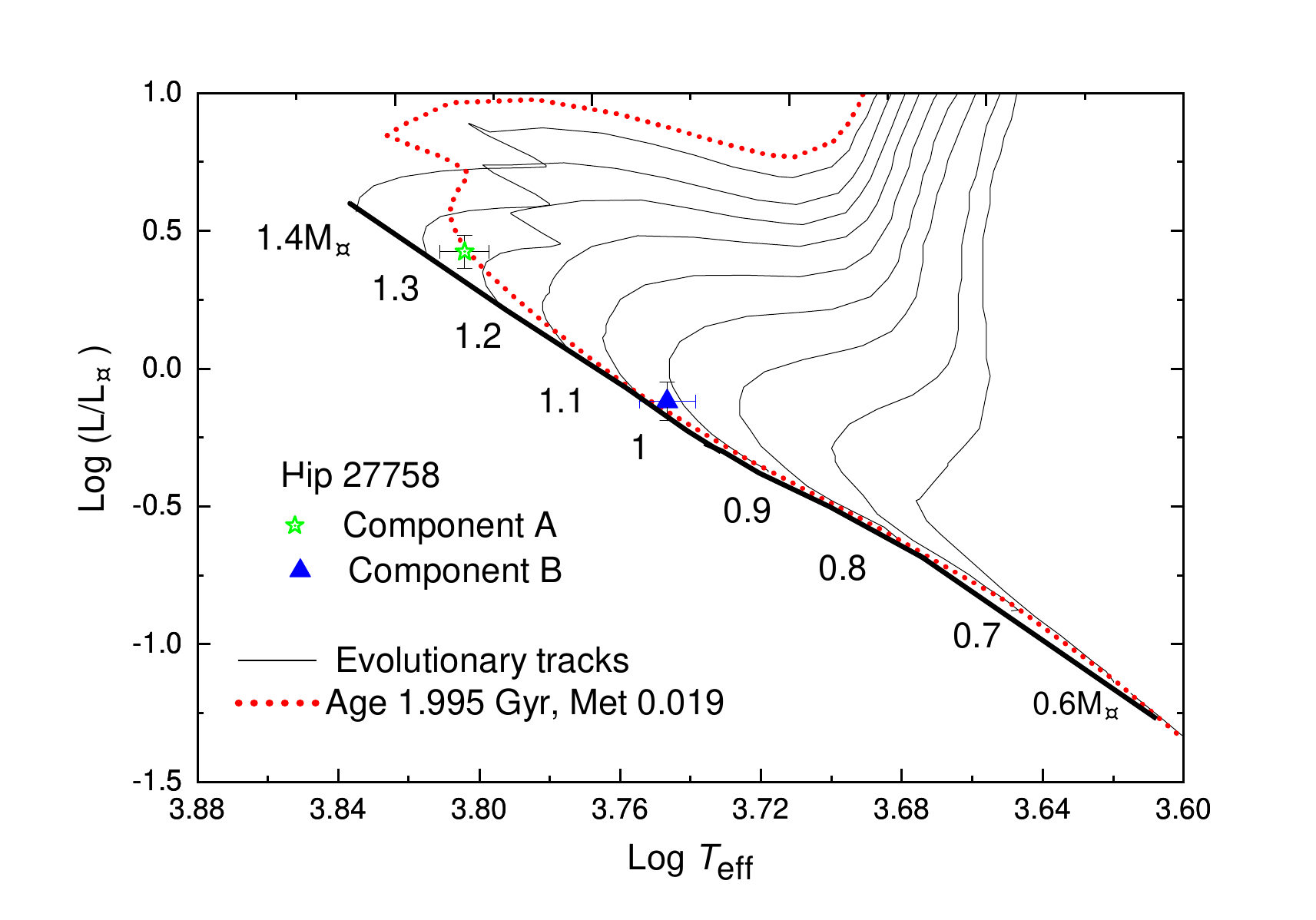}
	\caption{The synthetic evolutionary tracks of \cite{2000yCat..41410371G} and isochrones tracks of \cite{2000A&AS..141..371G} of both components of the system on the H-R diagram.}
	\label{f2}
\end{figure}

Fig. \ref{f2} shows the spectrophotometric stellar masses of HD\,39438, which are determined by using Al-Wardat's complex method for analyzing BMSSs based on the synthetic evolutionary tracks of \cite{2000yCat..41410371G} and fundamental stellar parameters of the system. These are found to be $1.24\pm0.11\,\rm M_\odot$ and  $0.98\pm0.09\,\rm M_\odot$ for the primary and secondary components of HD\,39438. According to \cite{2017AJ....154..110T}, the total mass of the system was $2.26 \mathcal M_\odot$ by using the spectral types, which is in keeping with our results ($2.22 \mathcal M_\odot$).

The total dynamical mass obtained by the orbital solutions of \cite{2010AJ....140..735M} ($2.56\pm0.32$) is consistent with the total mass achieved in this study within error margins, while there is no agreement between the results of Al-Wardat's method and dynamical mass obtained utilizing \cite{2014AJ....147..123T}'s orbit. However, \cite{2017AJ....154..110T} revised the orbit and presented new orbital parameters, the new orbital solution was
graded two (G 2=good), which is more accurate than the previous one. In his study, \cite{2017AJ....154..110T} presented
a new parallax as  $\pi_{dyn}=17.6$ mas depending on the new orbital solution. This further supports
our conclusion that the measured parallax for this system is not accurate enough and needs to be revised
by observations.

In our analysis, we used the dynamical parallax and the good orbital solution by \cite{2017AJ....154..110T} ($P=11.963\pm0.036$ yr and $a=0.1207\pm0.0007$ arcsec) to calculate the dynamical  mass sum as  $\Sigma \mathcal{M} =2.25\pm0.05 \mathcal{M}_{\odot}$, which is well in line with the spectrophotometric mass sum ($\Sigma \mathcal{M} =2.22\mathcal{M}_{\odot}$)
using Al-Wardat's method. In our case, we say that the suggested dynamical parallax should be slightly larger than the dynamical parallax of \cite{2017AJ....154..110T} based on our results. As a results, we used good orbital solution of \cite{2017AJ....154..110T} and our spectrophotometric  mass sum ($\Sigma \mathcal{M} =2.22\mathcal{M}_{\odot}$) to compute the new dynamical parallax as $\pi_{dyn}= 17.689\pm0.03$ mas, which is the closest estimate to the dynamical parallax of \cite{2017AJ....154..110T}. As a result, we expect that Gaia will perform and give good improvement in terms of the trigonometric
parallax in the near future.

Fig. \ref{f2} shows the positions of both components on the isochrones tracks of \cite{2000A&AS..141..371G}. In that case, we can see that the metallicity of HD\,39438 is [Z=0.019, Y=0.27], as shown in Fig. \ref{f2}. 

 Based on Fig. \ref{f2}, the age of the system is found to be 1.995 Gyr. The combined metallicity of the system was 0.015 based on the observed data \citep{2016ApJ...826..171G}, which corresponds
well with the synthetic metallicity of 0.019, as shown in Fig. \ref{f2}.

\section{Conclusions}\label{section6}
We have presented the fundamental stellar parameters of the close binary system, HD\,39438 using Al-Wardat's method for analyzing BMSSs. The method implements Kurucz's plane parallel model atmospheres
to construct synthetic SEDs for both components of the system. It then combines the
results of the spectroscopic analysis with the photometric analysis and then compares them with the observed ones to construct the best synthetic SEDs for the combined system. The best match between the synthetic and observed magnitudes and colour indices of the system for various photometrical systems, including Johnson: $U$, $B$, $ V$, $R$, $U-B$, $B-V$, $V-R$; Str\"{o}mgren: $u$, $v$, $b$,
$y$, $u-v$, $v-b$, $b-y$ and Tycho: $B_{T}$, $ V_{T}$, $B_{T}-V_{T}$ is showcased. 

The results shows that HD\,39438  consists of two main sequence stars; a 1.24 solar mass
with F5.5V and a 0.98 solar mass with G8V, both have the same age around 2 Gyr. We revised the dynamical parallax of the system, which is estimated to be $16.689\pm0.03$ mas.

\begin{acknowledgements}

 This study utilized several resources and tools, including SAO/NASA, the SIMBAD database, the Fourth Catalog of Interferometric Measurements of Binary Stars, IPAC data systems, the ORBIT code and the CHORIZOS code for photometric and spectrophotometric data analysis, and codes of Al-Wardat's method for
	analyzing binary and multiple stellar systems (BMSSs).  

\end{acknowledgements}

\bibliographystyle{raa}
\bibliography{references}

\begin{thebibliography}{56}
\providecommand\natexlab[1]{#1}
\providecommand\JournalTitle[1]{#1}

\bibitem[{Al-Wardat}(2002)]{2002BSAO...53...51A}
{Al-Wardat}, M.~A. 2002, Bull.~Special Astrophys.~Obs., 53, 51

\bibitem[{Al-Wardat}(2003)]{2003BSAO...55...18A}
{Al-Wardat}, M.~A. 2003, Bulletin of the Special Astrophysics Observatory, 55,
  18

\bibitem[{Al-Wardat}(2007)]{2007AN....328...63A}
{Al-Wardat}, M.~A. 2007, Astronomische Nachrichten, 328, 63

\bibitem[{Al-Wardat}(2009)]{2009AN....330..385A}
{Al-Wardat}, M.~A. 2009, Astronomische Nachrichten, 330, 385

\bibitem[{Al-Wardat}(2012)]{2012PASA...29..523A}
{Al-Wardat}, M.~A. 2012, \pasa, 29, 523

\bibitem[{Al-Wardat}(2014)]{2014AstBu..69..454A}
{Al-Wardat}, M.~A. 2014, Astrophysical Bulletin, 69, 454

\bibitem[{Al-Wardat} {et~al.}(2014{\natexlab{a}})]{2014AstBu..69...58A}
{Al-Wardat}, M.~A., {Balega}, Y.~Y., {Leushin}, V.~V., {et~al.}
  2014{\natexlab{a}}, Astrophysical Bulletin, 69, 58

\bibitem[{Al-Wardat} {et~al.}(2014{\natexlab{b}})]{2014AstBu..69..198A}
{Al-Wardat}, M.~A., {Balega}, Y.~Y., {Leushin}, V.~V., {et~al.}
  2014{\natexlab{b}}, Astrophysical Bulletin, 69, 198

\bibitem[{Al-Wardat} {et~al.}(2017)]{2017AstBu..72...24A}
{Al-Wardat}, M.~A., {Docobo}, J.~A., {Abushattal}, A.~A., \& {Campo}, P.~P.
  2017, Astrophysical Bulletin, 72, 24

\bibitem[{Al-Wardat} {et~al.}(2016)]{2016RAA....16..166A}
{Al-Wardat}, M.~A., {El-Mahameed}, M.~H., {Yusuf}, N.~A., {Khasawneh}, A.~M.,
  \& {Masda}, S.~G. 2016, Research in Astronomy and Astrophysics, 16, 166

\bibitem[{Al-Wardat} {et~al.}(2021{\natexlab{a}})]{2021PASA...38....2A}
{Al-Wardat}, M.~A., {Hussein}, A.~M., {Al-Naimiy}, H.~M., \& {Barstow}, M.~A.
  2021{\natexlab{a}}, \pasa, 38, e002

\bibitem[{Al-Wardat} {et~al.}(2014{\natexlab{c}})]{2014PASA...31....5A}
{Al-Wardat}, M.~A., {Widyan}, H.~S., \& {Al-thyabat}, A. 2014{\natexlab{c}},
  \pasa, 31, e005

\bibitem[{Al-Wardat} {et~al.}(2021{\natexlab{b}})]{2021RAA....21..161A}
{Al-Wardat}, M.~A., {Abu-Alrob}, E., {Hussein}, A.~M., {et~al.}
  2021{\natexlab{b}}, Research in Astronomy and Astrophysics, 21, 161

\bibitem[{Balega} {et~al.}(2002{\natexlab{a}})]{2002AA...385...87B}
{Balega}, I.~I., {Balega}, Y.~Y., {Hofmann}, K.-H., {et~al.}
  2002{\natexlab{a}}, \aap, 385, 87

\bibitem[{Balega} {et~al.}(2002{\natexlab{b}})]{2002AL...28..773B}
{Balega}, Y.~Y., {Tokovinin}, A.~A., {Pluzhnik}, E.~A., \& {Weigelt}, G.
  2002{\natexlab{b}}, Astronomy Letters, 28, 773

\bibitem[{Duquennoy} {et~al.}(1991)]{1991A&AS...88..281D}
{Duquennoy}, A., {Mayor}, M., \& {Halbwachs}, J.-L. 1991, \aaps, 88, 281

\bibitem[{ESA}(1997)]{1997yCat.1239....0E}
{ESA}. 1997, {The Hipparcos and Tycho Catalogues (ESA)}

\bibitem[{Gaia Collaboration}(2020)]{2020yCat.1350....0G}
{Gaia Collaboration}. 2020, VizieR Online Data Catalog, I/350

\bibitem[{Gaia Collaboration}(2022)]{2022yCat.1355....0G}
{Gaia Collaboration}. 2022, VizieR Online Data Catalog, I/355

\bibitem[{Gaia Collaboration} {et~al.}(2016)]{2016A&A...595A...1G}
{Gaia Collaboration}, {Prusti}, T., {de Bruijne}, J.~H.~J., {et~al.} 2016,
  \aap, 595, A1

\bibitem[{G{\'a}sp{\'a}r} {et~al.}(2016)]{2016ApJ...826..171G}
{G{\'a}sp{\'a}r}, A., {Rieke}, G.~H., \& {Ballering}, N. 2016, \apj, 826, 171

\bibitem[{Girardi} {et~al.}(2000{\natexlab{a}})]{2000A&AS..141..371G}
{Girardi}, L., {Bressan}, A., {Bertelli}, G., \& {Chiosi}, C.
  2000{\natexlab{a}}, \aaps, 141, 371

\bibitem[{Girardi} {et~al.}(2000{\natexlab{b}})]{2000yCat..41410371G}
{Girardi}, L., {Bressan}, A., {Bertelli}, G., \& {Chiosi}, C.
  2000{\natexlab{b}}, VizieR Online Data Catalog, 414, 10371

\bibitem[{Gray}(2005)]{2005oasp.book.....G}
{Gray}, D.~F. 2005, {The Observation and Analysis of Stellar Photospheres}, 505

\bibitem[{Hartkopf} {et~al.}(2012)]{2012AJ....143...42H}
{Hartkopf}, W.~I., {Tokovinin}, A., \& {Mason}, B.~D. 2012, \aj, 143, 42

\bibitem[{Hauck} \& {Mermilliod}(1998)]{1998A&AS..129..431H}
{Hauck}, B., \& {Mermilliod}, M. 1998, \aaps, 129, 431

\bibitem[{H{\o}g} {et~al.}(2000)]{2000A&A...355L..27H}
{H{\o}g}, E., {Fabricius}, C., {Makarov}, V.~V., {et~al.} 2000, \aap, 355, L27

\bibitem[{Horch} {et~al.}(2010)]{2010AJ....139..205H}
{Horch}, E.~P., {Falta}, D., {Anderson}, L.~M., {et~al.} 2010, \aj, 139, 205

\bibitem[{Horch} {et~al.}(2008)]{2008AJ....136..312H}
{Horch}, E.~P., {van Altena}, W.~F., {Cyr}, Jr., W.~M., {et~al.} 2008, \aj,
  136, 312

\bibitem[{Houk} \& {Swift}(1999)]{1999MSS...C05....0H}
{Houk}, N., \& {Swift}, C. 1999, Michigan Spectral Survey, 5, 0

\bibitem[{K{\"o}hler}(2014)]{2014CoSka..43..229K}
{K{\"o}hler}, R. 2014, Contributions of the Astronomical Observatory Skalnate
  Pleso, 43, 229

\bibitem[{Kurucz}(1994)]{1994KurCD..19.....K}
{Kurucz}, R. 1994, Solar abundance model atmospheres for 0,1,2,4,8 km/s.~Kurucz
  CD-ROM No.~19.~ Cambridge, Mass.: Smithsonian Astrophysical Observatory,
  1994., 19

\bibitem[{Lallement} {et~al.}(2014)]{2014A&A...561A..91L}
{Lallement}, R., {Vergely}, J.~L., {Valette}, B., {et~al.} 2014, \aap, 561, A91

\bibitem[{Lang}(1992)]{1992adps.book.....L}
{Lang}, K.~R. 1992, {Astrophysical Data I. Planets and Stars.}, 133

\bibitem[{Lucy}(2018)]{2018A&A...618A.100L}
{Lucy}, L.~B. 2018, \aap, 618, A100

\bibitem[{Ma{\'{\i}}z Apell{\'a}niz}(2007)]{2007ASPC..364..227M}
{Ma{\'{\i}}z Apell{\'a}niz}, J. 2007, in Astronomical Society of the Pacific
  Conference Series, Vol. 364, The Future of Photometric, Spectrophotometric
  and Polarimetric Standardization, ed. C.~{Sterken} (San Francisco:
  Astronomical Society of the Pacific), 227

\bibitem[Masda \& Al-Wardat(2023)]{MASDA2023}
Masda, S., \& Al-Wardat, M. 2023, Advances in Space Research

\bibitem[{Masda} {et~al.}(2016)]{2016RAA....16..112M}
{Masda}, S.~G., {Al-Wardat}, M.~A., {Neuh{\"a}user}, R., \& {Al-Naimiy}, H.~M.
  2016, Research in Astronomy and Astrophysics, 16, 112

\bibitem[{Masda} {et~al.}(2018{\natexlab{a}})]{2018RAA....18...72M}
{Masda}, S.~G., {Al-Wardat}, M.~A., \& {Pathan}, J. K. M.~K.
  2018{\natexlab{a}}, Research in Astronomy and Astrophysics, 18, 072

\bibitem[{Masda} {et~al.}(2018{\natexlab{b}})]{2018JApA...39...58M}
{Masda}, S.~G., {Al-Wardat}, M.~A., \& {Pathan}, J.~M. 2018{\natexlab{b}},
  Journal of Astrophysics and Astronomy, 39, 58

\bibitem[{Masda} {et~al.}(2019{\natexlab{a}})]{2019RAA....19..105M}
{Masda}, S.~G., {Al-Wardat}, M.~A., \& {Pathan}, J.~M. 2019{\natexlab{a}},
  Research in Astronomy and Astrophysics, 19, 105

\bibitem[{Masda} {et~al.}(2019{\natexlab{b}})]{2019AstBu..74..464M}
{Masda}, S.~G., {Docobo}, J.~A., {Hussein}, A.~M., {et~al.} 2019{\natexlab{b}},
  Astrophysical Bulletin, 74, 464

\bibitem[{Masda} {et~al.}(2021)]{2021AIPC.2335i0002M}
{Masda}, S.~G., {Khan}, A.~R., \& {Pathan}, J.~M. 2021, in American Institute
  of Physics Conference Series, Vol. 2335, American Institute of Physics
  Conference Series, 090002

\bibitem[{Mason} {et~al.}(2010)]{2010AJ....140..735M}
{Mason}, B.~D., {Hartkopf}, W.~I., \& {Tokovinin}, A. 2010, \aj, 140, 735

\bibitem[{Pluzhnik}(2005)]{2005AA...431..587P}
{Pluzhnik}, E.~A. 2005, \aap, 431, 587

\bibitem[{Roberts}(2011)]{2011MNRAS.413.1200R}
{Roberts}, Jr., L.~C. 2011, \mnras, 413, 1200

\bibitem[{Roberts} {et~al.}(2005)]{2005AJ....130.2262R}
{Roberts}, Jr., L.~C., {Turner}, N.~H., {Bradford}, L.~W., {et~al.} 2005, \aj,
  130, 2262

\bibitem[{Tokovinin}(2016)]{2016AJ....152..138T}
{Tokovinin}, A. 2016, \aj, 152, 138

\bibitem[{Tokovinin}(2017)]{2017AJ....154..110T}
{Tokovinin}, A. 2017, \aj, 154, 110

\bibitem[{Tokovinin} {et~al.}(2000)]{Tokovinin2000}
{Tokovinin}, A.~A., {Balega}, Y.~Y., {Hofmann}, K.~H., \& {Weigelt}, G. 2000,
  Astronomy Letters, 26, 668

\bibitem[{Tokovinin} {et~al.}(2010)]{2010AJ....139..743T}
{Tokovinin}, A., {Mason}, B.~D., \& {Hartkopf}, W.~I. 2010, \aj, 139, 743

\bibitem[{Tokovinin} {et~al.}(2014)]{2014AJ....147..123T}
{Tokovinin}, A., {Mason}, B.~D., \& {Hartkopf}, W.~I. 2014, \aj, 147, 123

\bibitem[{Tokovinin} {et~al.}(2015)]{2015AJ....150...50T}
{Tokovinin}, A., {Mason}, B.~D., {Hartkopf}, W.~I., {Mendez}, R.~A., \&
  {Horch}, E.~P. 2015, \aj, 150, 50

\bibitem[{van Leeuwen}(2007)]{2007A&A...474..653V}
{van Leeuwen}, F. 2007, \aap, 474, 653

\bibitem[{Widyan} \& {Aljboor}(2021)]{2021RAA....21..110W}
{Widyan}, H., \& {Aljboor}, H. 2021, Research in Astronomy and Astrophysics,
  21, 110

\bibitem[{Yousef} {et~al.}(2021)]{2021RAA....21..114Y}
{Yousef}, Z.~T., {Annuar}, A., {Hussein}, A.~M., {et~al.} 2021, Research in
  Astronomy and Astrophysics, 21, 114

\end{thebibliography}


\begin{thebibliography}{10}

\bibitem{2007AN....328...63A}
M.~A. {Al-Wardat}.
\newblock {Model atmosphere parameters of the binary systems COU1289 and
  COU1291}.
\newblock {\em Astronomische Nachrichten}, 328:63--67, January 2007.

\bibitem{2004AJ....127.1727H}
E.~P. {Horch}, R.~D. {Meyer}, and W.~F. {van Altena}.
\newblock {Speckle Observations of Binary Stars with the WIYN Telescope. IV.
  Differential Photometry}.
\newblock {\em \aj}, 127:1727--1735, March 2004.

\bibitem{2008AJ....136..312H}
E.~P. {Horch}, W.~F. {van Altena}, W.~M. {Cyr}, Jr., L.~{Kinsman-Smith},
  A.~{Srivastava}, and J.~{Zhou}.
\newblock {Charge-Coupled Device Speckle Observations of Binary Stars with the
  WIYN Telescope. V. Measures During 2001-2006}.
\newblock {\em \aj}, 136:312--322, July 2008.

\bibitem{2011AJ....141..180H}
E.~P. {Horch}, W.~F. {van Altena}, S.~B. {Howell}, W.~H. {Sherry}, and D.~R.
  {Ciardi}.
\newblock {Observations of Binary Stars with the Differential Speckle Survey
  Instrument. III. Measures below the Diffraction Limit of the WIYN Telescope}.
\newblock {\em \aj}, 141:180, June 2011.

\bibitem{2010AJ....139..205H}
E.~P. {Horch}, D.~{Falta}, L.~M. {Anderson}, M.~D. {DeSousa}, C.~M. {Miniter},
  T.~{Ahmed}, and W.~F. {van Altena}.
\newblock {CCD Speckle Observations of Binary Stars with the WIYN Telescope.
  VI. Measures During 2007-2008}.
\newblock {\em \aj}, 139:205--215, January 2010.

\bibitem{2012AJ....143...10H}
E.~P. {Horch}, L.~A.~P. {Bahi}, J.~R. {Gaulin}, S.~B. {Howell}, W.~H. {Sherry},
  R.~{Baena Gall{\'e}}, and W.~F. {van Altena}.
\newblock {Speckle Observations of Binary Stars with the WIYN Telescope. VII.
  Measures during 2008-2009}.
\newblock {\em \aj}, 143:10, January 2012.

\bibitem{2011AJ....141...45H}
E.~P. {Horch}, S.~C. {Gomez}, W.~H. {Sherry}, S.~B. {Howell}, D.~R. {Ciardi},
  L.~M. {Anderson}, and W.~F. {van Altena}.
\newblock {Observations of Binary Stars with the Differential Speckle Survey
  Instrument. II. Hipparcos Stars Observed in 2010 January and June}.
\newblock {\em \aj}, 141:45, February 2011.

\bibitem{2006A&A...448..703B}
I.~I. {Balega}, Y.~Y. {Balega}, K.-H. {Hofmann}, E.~V. {Malogolovets},
  D.~{Schertl}, Z.~U. {Shkhagosheva}, and G.~{Weigelt}.
\newblock {Orbits of new Hipparcos binaries. II}.
\newblock {\em \aap}, 448:703--707, March 2006.

\bibitem{1992adps.book.....L}
K.~R. {Lang}.
\newblock {\em {Astrophysical Data I. Planets and Stars.}}
\newblock 1992.

\bibitem{2005oasp.book.....G}
D.~F. {Gray}.
\newblock {\em {The Observation and Analysis of Stellar Photospheres}}.
\newblock September 2005.

\bibitem{1994KurCD..19.....K}
R.~{Kurucz}.
\newblock {Solar abundance model atmospheres for 0,1,2,4,8 km/s.}
\newblock {\em Solar abundance model atmospheres for 0,1,2,4,8 km/s.~Kurucz
  CD-ROM No.~19.~ Cambridge, Mass.: Smithsonian Astrophysical Observatory,
  1994.}, 19, 1994.

\bibitem{2007ASPC..364..227M}
J.~{Ma{\'{\i}}z Apell{\'a}niz}.
\newblock {A Uniform Set of Optical/NIR Photometric Zero Points to be Used with
  CHORIZOS}.
\newblock In C.~{Sterken}, editor, {\em The Future of Photometric,
  Spectrophotometric and Polarimetric Standardization}, volume 364 of {\em
  Astronomical Society of the Pacific Conference Series}, pages 227--236. San
  Francisco: Astronomical Society of the Pacific, April 2007.

\bibitem{2012PASA...29..523A}
M.~A. {Al-Wardat}.
\newblock {Physical Parameters of the Visually Close Binary Systems Hip70973
  and Hip72479}.
\newblock {\em \pasa}, 29:523--528, June 2012.

\bibitem{2000yCat..41410371G}
L.~{Girardi}, A.~{Bressan}, G.~{Bertelli}, and C.~{Chiosi}.
\newblock {VizieR Online Data Catalog: Low-mass stars evolutionary tracks
  {\amp} isochrones (Girardi+, 2000)}.
\newblock {\em VizieR Online Data Catalog}, 414:10371, November 2000.

\bibitem{2007A&A...474..653V}
F.~{van Leeuwen}.
\newblock {Validation of the new Hipparcos reduction}.
\newblock {\em \aap}, 474:653--664, November 2007.

\bibitem{1992ASPC...32..573T}
A.~{Tokovinin}.
\newblock {Speckle Spectroscopic Studies of Late-Type Stars}.
\newblock In H.~A. {McAlister} and W.~I. {Hartkopf}, editors, {\em IAU Colloq.
  135: Complementary Approaches to Double and Multiple Star Research},
  volume~32 of {\em Astronomical Society of the Pacific Conference Series},
  page 573, 1992.

\bibitem{2000A&AS..141..371G}
L.~{Girardi}, A.~{Bressan}, G.~{Bertelli}, and C.~{Chiosi}.
\newblock {Evolutionary tracks and isochrones for low- and intermediate-mass
  stars: From 0.15 to 7 M$_{sun}$, and from Z=0.0004 to 0.03}.
\newblock {\em \aaps}, 141:371--383, February 2000.

\end{thebibliography}

\end{document}